\documentclass[12pt]{article}%
\usepackage[nosort]{cite}
\usepackage[usenames, dvipsnames]{xcolor}
\usepackage{graphicx}
\usepackage{multicol}
\usepackage{amsfonts}
\usepackage{amssymb}
\usepackage{amsmath}
\usepackage{torres}
\usepackage{afterpage}
\usepackage{braket}
\usepackage{setspace}
\usepackage{verbatim}
\usepackage{color}
\usepackage{longtable}
\usepackage{float}
\usepackage{subcaption}
\usepackage{epsfig}
\usepackage{enumerate}
\usepackage{epstopdf}
\usepackage[enableskew, vcentermath]{youngtab}
\usepackage{adjustbox}
\usepackage{multirow}
\usepackage{tikz}
\usepackage[margin=1in]{geometry}
\usepackage{titletoc}
\usepackage[percent]{overpic}
\usepackage{gensymb}
\usepackage{bbm}
\usepackage{mathtools}
\usepackage{tikz-cd}%
\providecommand{\U}[1]{\protect\rule{.1in}{.1in}}
\newsavebox{\mysavebox}

\usetikzlibrary{decorations.markings}

\numberwithin{equation}{section}

\usetikzlibrary{chains}
\allowdisplaybreaks
\tikzset{node distance=2em, ch/.style={circle,draw,on chain,inner sep=2pt},chj/.style={ch,join},every path/.style={shorten >=4pt,shorten <=4pt},line width=1pt,baseline=-1ex}

\tikzstyle{startstop} = [rectangle, rounded corners, minimum width=3cm, minimum height=1cm,text centered, draw=black, fill=blue!10]
\tikzstyle{startstop} = [rectangle, rounded corners, minimum width=3cm, minimum height=1cm,text centered, draw=black, fill=blue!10]
\tikzstyle{io} = [trapezium, trapezium left angle=70, trapezium right angle=110, minimum width=3cm, minimum height=1cm, text centered, draw=black, fill=blue!30]
\tikzstyle{process} = [rectangle, minimum width=3cm, minimum height=1cm, text centered, draw=black, fill=orange!30]
\tikzstyle{decision} = [diamond, minimum width=3cm, minimum height=1cm, text centered, draw=black, fill=green!30]
\tikzstyle{arrow} = [thick,->,>=stealth]
\tikzset{->-/.style={decoration={
  markings,
  mark=at position #1 with {\arrow[scale=2.4]{>}}},postaction={decorate}}}
\usetikzlibrary{arrows}
\usetikzlibrary{arrows.meta}
\usetikzlibrary{arrows,decorations.pathmorphing}
\usetikzlibrary{shapes.geometric,calc,arrows, positioning,shapes.misc,decorations.markings}
\tikzset{
  big arrow/.style={
    decoration={markings,mark=at position 1 with {\arrow[scale=2,#1]{>}}},
    postaction={decorate},
    shorten >=0.4pt},
  big arrow/.default=black}
\pgfdeclarelayer{edgelayer}
\pgfdeclarelayer{nodelayer}
\pgfsetlayers{edgelayer,nodelayer,main}
\tikzstyle{none}=[inner sep=0pt]

\tikzstyle{NodeCross}=[draw, shape=circle, cross out, inner sep=0pt, minimum size=10pt,line width=0.25mm]
\tikzstyle{Circle}=[draw, shape=circle, black, fill=black, inner sep=0pt, minimum size=4pt]
\tikzstyle{circle}=[draw, shape=circle, black, fill=black, inner sep=0pt, minimum size=16pt]
\tikzstyle{CircleRed}=[draw, shape=circle, fill={rgb,255: red,191; green,0; blue,0}, inner sep=0pt, minimum size=8pt]
\tikzstyle{Star}=[draw, shape=star, fill=red, star points=8, inner sep=0pt, minimum size=12pt]
\tikzstyle{CircleBlue}=[draw, shape=circle, fill=blue, inner sep=0pt, minimum size=8pt]
\tikzstyle{CirclePurple}=[draw, shape=circle, fill={rgb,255: red,191; green,0; blue,191}, inner sep=0pt, minimum size=8pt]
\tikzstyle{EmptyCircle}=[draw, shape=circle, inner sep=0pt, minimum size=4pt]
\tikzstyle{GreenCircle}=[draw, shape=circle,  fill={rgb,255: red,80; green,200; blue,120}, inner sep=0pt, minimum size=8pt]
\tikzstyle{BrownCircle}=[draw, shape=circle,  fill={rgb,255: red,115; green,115; blue,115}, inner sep=0pt, minimum size=8pt]
\tikzstyle{CirclePurpleSmall}=[draw, shape=circle, fill={rgb,255: red,191; green,0; blue,191}, inner sep=0pt, minimum size=4pt]

\tikzstyle{DashedLine}=[-, densely dashed, line width=0.25mm]
\tikzstyle{DottedLine}=[-, dotted, line width=0.25mm]
\tikzstyle{ThickLine}=[-, line width=0.25mm]
\tikzstyle{ArrowLineRight}=[-, -{Stealth[scale=1.75]}, line width=0.15mm, scale=5]
\tikzstyle{RedLine}=[-, draw={rgb,255: red,191; green,0; blue,0}, fill=none, line width=0.5mm]
\tikzstyle{DashedLineThin}=[-, densely dashed, line width=0.125mm, fill=none, draw=black]
\tikzstyle{ArrowLineRed}=[-, draw={rgb,255: red,191; green,0; blue,0}, -{Stealth[scale=1.75]}, line width=0.15mm, scale=5]
\tikzstyle{BlueLine}=[-, draw={rgb,255: red,0; green,0; blue,191}, fill=none, line width=0.5mm]
\tikzstyle{DashedRed}=[-, densely dashed, draw={rgb,255: red,191; green,0; blue,0}, fill=none, line width=0.5mm]
\tikzstyle{DottedRed}=[-, dotted, draw={rgb,255: red,191; green,0; blue,0}, fill=none, dotted, line width=0.5mm]
\tikzstyle{BlueDottedLight}=[-, dotted, draw={rgb,255: red,0; green,0; blue,191}, fill=none, line width=0.5mm]
\tikzstyle{ArrowLinePurple}=[-, draw={rgb,255: red,191; green,0; blue,191}, -{Stealth[scale=1.75]}, line width=0.15mm, scale=5]
\tikzstyle{DashedLineGreen}=[-, densely dashed, draw={rgb,255: red,74; green,103; blue,65}, line width=0.25mm]
\tikzstyle{LineGreen}=[-, draw={rgb,255: red, 74; green,200; blue,65}, line width=0.5mm]

\tikzset{snake it/.style={decorate, decoration=snake}}

\tikzstyle{PurpleLine}=[-, draw={rgb,255: red,191; green,0; blue,191}, fill=none, line width=0.5mm]
\tikzstyle{BrownLine}=[-, draw={rgb,255: red,115; green,115; blue,115}, fill=none, line width=0.5mm]

\usetikzlibrary{decorations.markings}
\makeatletter \@addtoreset{equation}{section} \makeatother

\colorlet{darkblue}{blue!70!black}
\colorlet{darkgreen}{green!70!black}

\usepackage[colorlinks=true,urlcolor=darkblue,linktocpage=true,linkcolor=darkblue,citecolor=darkblue]{hyperref}
\hypersetup{
colorlinks=true,
linkcolor=blue,
citecolor=magenta,
}
\begin{document}

%% Report number
\vspace*{-2cm}
\begin{flushright}
{\tt CERN-TH-2024-054}\\
\end{flushright}

\date{May 2024}

\title{Giving a $KO$ to 8D Gauge Anomalies}
%\title{On the Topological Green-Schwarz Mechanism \\[5mm] and K-theoretic Symmetries}

\institution{CERN}{\centerline{${}^{1}$CERN Theory Department, CH-1211 Geneva, Switzerland}}

%\institution{IPMU}{\centerline{${}^{8}$Kavli IPMU, University of Tokyo, Kashiwa, Chiba 277-8583, Japan}}

\authors{Ethan Torres\worksat{\CERN}}

\abstract{In \cite{Garcia-Etxebarria:2017crf}, it was found that the system of $k$ D7-branes probing an $O7^+$-plane suffers from an $\mathfrak{sp}(k)$ gauge anomaly when $k>1$. These authors then conjectured that this 8D $\mathcal{N}=1$ gauge theory couples to an 8D topological field theory (TFT) such that the total system is anomaly-free, thus acting as a ``topological" Green-Schwarz mechanism. In this note, we construct such an 8D TFT and show that it indeed cancels the gauge anomaly. The key step is to engineer the relevant topological operators from D3-branes and fluxbranes placed infinitely far away from the stack of 7-branes. Such symmetry operators have topological vector bundles defined on them whose $KO/KSp$-homology classes play a role in the anomaly cancellation.}

\maketitle

\setcounter{tocdepth}{2}

%\enlargethispage{\baselineskip}

\newpage

\section{Introduction}
Over the past few years there has been much progress on understanding generalized/categorical global symmetries of quantum field theories (QFTs). For recent reviews see \cite{Cordova:2022ruw, Schafer-Nameki:2023jdn, Bhardwaj:2023kri, Luo:2023ive, Shao:2023gho}. As with textbook global symmetries, the data of 't Hooft anomalies for these more general symmetries can be especially powerful in extracting non-perturbative information of a QFT. One construction that has been especially useful in this program has been that of a Symmetry topological field theory (SymTFT). For a given $D$-dimensional QFT of interest, its SymTFT is a $(D+1)$-dimensional TFT which carries all of the information of the global symmetries and anomalies of the QFT\footnote{This statement is most precise for \textit{discrete} categorical symmetries, although much recent progress has been made on incorporating continuous symmetries into this framework see \cite{Brennan:2024fgj,Antinucci:2024zjp,Bonetti:2024cjk, Apruzzi:2024htg}. }.

Such a perspective is particularly natural when a QFT is realized in string theory whereby there is typically a notion of ``radial direction" in the extra dimensions of the string construction pointing away from where the QFT is localized. This radial direction plays the role of the extra dimension of the SymTFT \cite{Apruzzi:2021nmk}. This leads to a powerful dictionary between string theory objects (possibly wrapped on cycles in the internal geometry) and field theory objects. For instance, branes wrapping the ``radial" direction give rise to local/extended operators which are charged under global symmetries whose topological operators are engineered from wrapping branes infinitely far away from the $D$-dimensional QFT degrees of freedom \cite{Heckman:2022muc}. Interesting properties of symmetry operators such as non-invertibility and higher-group structures are then the direct result of string theory input such as Wess-Zumino terms on branes and Hanany-Witten transitions.

One may also entertain the possibility that string theory can predict new features of generalized global symmetries rather than reproducing properties that have previously appeared in the field theory literature. As remarked in the conclusions of \cite{Heckman:2022muc}, such applications are expected to follow from the fact that string theory branes possess a bevy of intricate structures\footnote{For instance, derived categories are known to classify D-branes in topological twisted string theories and reflect BPS objects in physical string theories.}. Indeed, if a $D$-dimensional QFT is engineered from D-branes/O-planes, then topological symmetry operators that arise from D-branes placed infinitely far away possess vector bundles localized on their worldvolume which persist even after integrating out massive bifundamental strings. Such symmetry operators are labeled by K-theory charges as mentioned in Appendix A of \cite{Heckman:2022xgu} (see also the recent \cite{Zhang:2024oas} which refers to these as ``K-theoretic symmetries").

In this note, we employ such symmetry operators to address a key issue raised in \cite{Garcia-Etxebarria:2017crf} concerning the system of $k$ D7-branes probing a $O7^+$-plane. This system engineers an 8D $\mathcal{N}=1$ $\mathfrak{sp}(k)$ gauge theory at low energies which, at face value, has a subtle $\mathbb{Z}_2$ valued gauge anomaly when $k>1$ due to the adjoint Majorana fermion. We will denote this gauge theory by $\mathcal{T}_{\mathfrak{sp}(k)}$. Recall that unlike 't Hooft anomalies, gauge anomalies spell an inconsistency in the theory which must be canceled. As mentioned in \cite{Garcia-Etxebarria:2017crf}, the structure of the 8D $\mathcal{N}=1$ supersymmetry is sufficiently constraining that this anomaly cannot be canceled by coupling to additional massless degrees of freedom, so these authors conjectured was that there is some 8D TFT that couples to the $\mathfrak{sp}(k)$ gauge theory which makes the total 8D theory anomaly free. This was naturally referred to as a topological Green-Schwarz mechanism\footnote{The original Green-Schwarz mechanism of \cite{Green:1984sg} also involved a topological coupling albeit between gapless (i.e. non-topological) excitations.}.

\begin{figure}
\centering
\includegraphics[scale=0.35, trim = {0cm 0cm 0cm 0cm}]{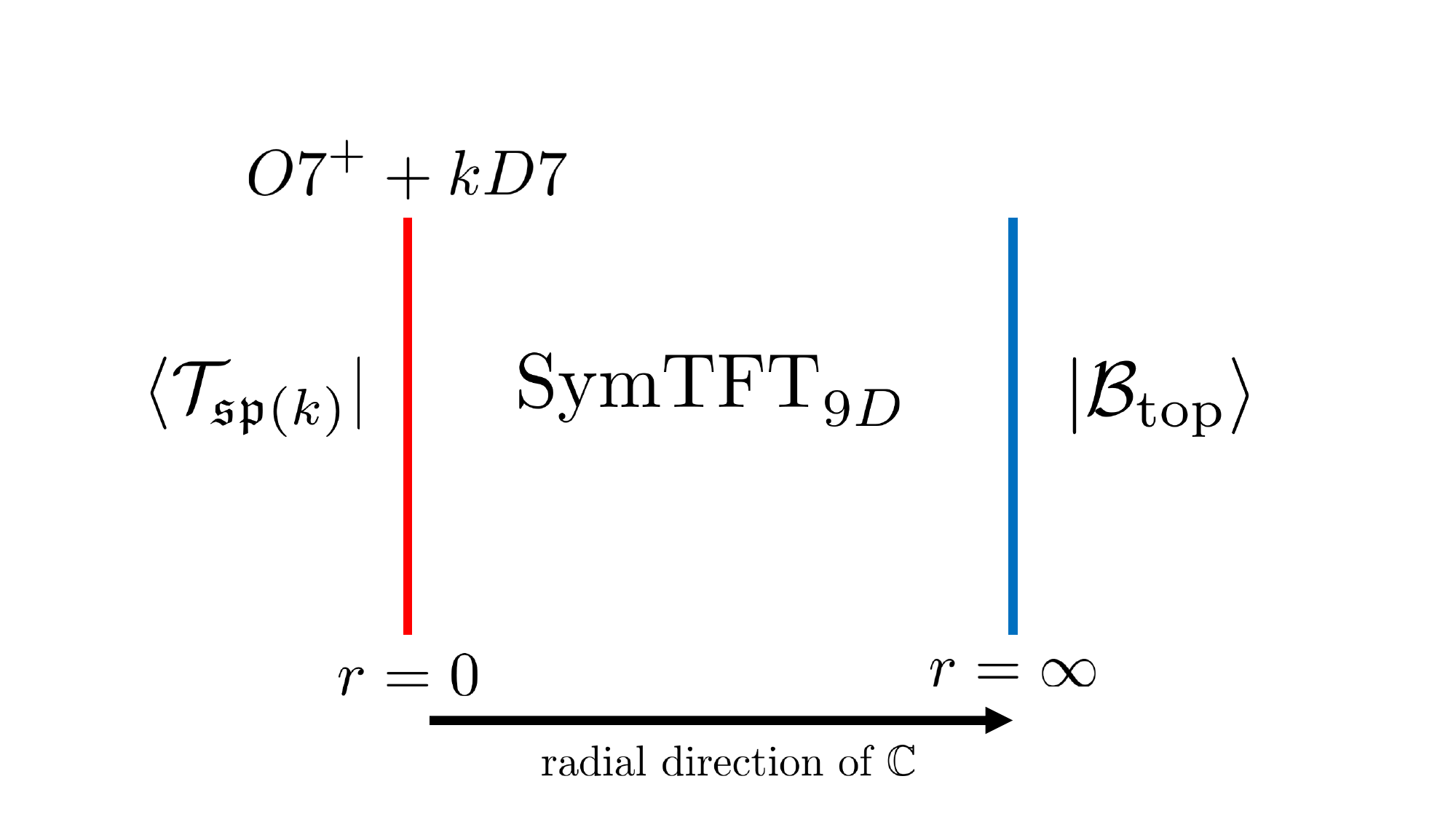}
\caption{We illustrate the 9D SymTFT setup for the $O7^++kD7$ system where the radial direction of the normal directions to the 7-brane stack, $\mathbb{C}$, is identified with the interval direction of the SymTFT. The SymTFT has two boundaries one gapless (where the gauge theory lives) and one gapped whose boundary states are respectively $\ket{\mathcal{T}_{\mathfrak{sp}(k)}}$ and $\ket{\mathcal{B}_{\mathrm{top}}}$. }
\label{fig:SymTFT9D}
\end{figure}

The goal of this note then will be to construct such a 8D TFT and show that its coupling to $\mathcal{T}_{\mathfrak{sp}(k)}$ results in an anomaly-free 8D theory. Our strategy, following the philosophy of \cite{Heckman:2022muc}, will be to engineer the topological operators for this 8D theory by placing branes infinitely far away from the stack of 7-branes\footnote{We will not be exhaustive in describing all of the operators/fusion algebras of the 8D TFT, but rather just those features relevant to cancelling the anomaly.}. More precisely, such branes will be realized as topological operators in the 9D SymTFT which is defined by compactifying IIB on the infinitely large asymptotic $S^1_\infty$ surrounding the 7-brane stack. One then arrives at an 8D theory by compactifying the interval direction\footnote{We have implicitly sent the 10D Newton's constant to zero before compactifying the interval direction which is necessary for the 9D SymTFT to be topological in the first place.}. For a field theory illustration of the 9D SymTFT see Figure \ref{fig:SymTFT9D}. In this setup, we have topological boundary conditions along $S^1_\infty$ which is specified by a boundary state $\ket{\mathcal{B}_\mathrm{top}}$ of the SymTFT, and a gapless boundary condition $\ket{\mathcal{T}_{\mathfrak{sp}(k)}}$ defining the presence of the 8D vectormultiplet. Compactifying the interval leads to an 8D gauge theory coupled to a TFT, or more schematically
\begin{equation}
    \mathcal{T}_{\mathfrak{sp}(k)}\; |\; \mathrm{SymTFT}_{9D}\; |\; \mathcal{B}_{\mathrm{top}} \quad \rightarrow \quad \mathcal{T}_{\mathfrak{sp}(k)}\otimes \mathrm{TFT}^{(\mathcal{B}_{\mathrm{top}})}_{8D},
\end{equation}
and its partition function (which depends on the boundary condition $\mathcal{B}_{\mathrm{top}}$) is given by
\begin{equation}
    Z^{(\mathcal{B}_{\mathrm{top}})}_{8D}=\braket{\mathcal{T}_{\mathfrak{sp}(k)}|\mathcal{B}_{\mathrm{top}}}.
\end{equation}
We emphasize that $Z^{(\mathcal{B}_{\mathrm{top}})}_{8D}$ does not depend on the interval size because the SymTFT is topological. As we will show, the field theoretic interpretation of the topological boundary condition $\mathcal{B}_{\mathrm{top}}$ is that it requires that one sums only over even charged $\mathfrak{sp}(k)$ instantons which forbids any anomalous gauge configurations. Such restrictions in the sum over topological sectors of a gauge theory is nothing new, see for instance \cite{Seiberg:2010qd,Tanizaki:2019rbk}. Also, the general procedure of specifying asymptotic boundary conditions in the extra dimensions of string theory is precisely how one fixes the global form of the gauge group\footnote{This has long been appreciated in holography, e.g. \cite{Witten:1998wy}.}, i.e. whether it we have $Sp(k)$ or $Sp(k)/\mathbb{Z}_2$ as the gauge group of our 7-brane system.

This note is organized as follows. In Section \ref{sec:rev} we review the $\mathfrak{sp}(k)$ gauge anomaly found in \cite{Garcia-Etxebarria:2017crf} and its formulation in terms of $KO$/$KSp$-theory. In the beginning of Section \ref{sec:topgs} we discuss a straightforward way to construct a topological coupling which cancels the anomaly of \cite{Garcia-Etxebarria:2017crf} which relies on some particular assumptions of the boundary conditions of RR-fields. In Section \ref{ssec:symopsflux} we employ the more careful of constructing the 9D SymTFT from reducing IIB branes and in Section \ref{ssec:halfd3s} we demonstrate how the anomaly cancels for any choice of boundary condition. Section \ref{sec:conc} contains our conclusions and outlook.

\section{Review of 8D $\mathcal{N}=1$ $\mathfrak{sp}(k)$ Gauge Anomaly}\label{sec:rev}

%\footnote{In the string construction, this is done by setting Dirichlet/Neumann boundary conditions for $B_2$/$B_6$ along the asymptotic spatial $S^1=\partial \mathbb{C}$ surrounding the 7-brane stack.}

We begin by reviewing the argument of \cite{Garcia-Etxebarria:2017crf} that an 8D $\mathcal{N}=1$ $\mathfrak{sp}(k)$ gauge theory is anomalous when $k>1$. This anomaly exists for either choice of the global form of the gauge group, i.e. $Sp(k)$ or $Sp(k)/\mathbb{Z}_2$, so we will just focus on the gauge algebra. We consider first the decomposition
\begin{equation}\label{eq:sp1subgroup}
\mathfrak{sp}(k)\supset \mathfrak{sp}(k-1)\oplus \mathfrak{sp}(1)
\end{equation}
and turn on a single instanton configuration in the $\mathfrak{sp}(1)$ subalgebra. This background breaks the gauge algebra to $\mathfrak{sp}(k-1)$ and the Dirac equation for the adjoint Majorana has zero-mode solutions localized on the 4-manifold for which the instanton is localized. From the decomposition of the adjoint representation under \eqref{eq:sp1subgroup}
\begin{equation}\label{eq:decompadj}
  \mathrm{Adj}(Sp(k))\rightarrow (\mathbf{2k-2})\otimes \mathbf{2}  + \mathbf{1}\otimes \mathrm{Adj}(Sp(1))+\mathrm{Adj}(Sp(k-1))\otimes \mathbf{1}
\end{equation}
we see that, in particular, there are fermion zero-modes in $(\mathbf{2k-2})$ of the unbroken gauge group. From index theory, the number of such zero-modes is twice the Dynkin index of $\mathbf{2}$, so we have a total of $2\times (1/2)=1$ 4D Weyl fermion in the $(\mathbf{2k-2})$ representation localized on the instanton. This 4D theory suffers from the usual Witten anomaly valued in $\pi_4(Sp(k-1))=\mathbb{Z}_2$ \cite{Witten:1982fp} which means that this background is not invariant under fermion parity $(-1)^F$. Since $(-1)^F$ must be preserved in any theory with a fermions, this leads to a phase ambiguity which makes the 8D theory itself is anomalous.

Such a phase ambiguity of the 8D partition function (before summing over $\mathfrak{sp}(k)$ gauge fields) can be packaged in terms of a TFT of one higher-dimension much in the same way that a SymTFT captures the 't Hooft anomalies of a field theory. Recall that the usual 4D Witten anomaly can be characterized by a 5D TFT whose partition function can either be $1$ or $-1$ and in particular evaluates to $-1$ on 5-manifolds of the form $\Sigma_4\times S^1$ where we have one $\mathfrak{sp}(k-1)$ instanton on $\Sigma_4$ and $S^1$ has periodic spin structure. This spin structure makes $\Sigma_4\times S^1$ a mapping torus with respect to an action by $(-1)^F$. This 5D TFT is simply the the mod-2 index of the Dirac operator extended to 5D, for details see \cite{Witten:1982fp} and the reviews of these statements in \cite{Garcia-Etxebarria:2017crf,Wang:2018qoy}. Altogether, this means that the partition function of the 9D TFT characterizing the 8D anomaly must evaluate to $-1$ on backgrounds of the form
\begin{equation}\label{eq:backgroundanom}
   N_9=S^1\times \Sigma_4\times M_4, \quad  \frac{1}{4}\int_{\Sigma_4} \mathrm{Tr}_{Sp(k)}F^2=2n_1+1, \quad \frac{1}{4}\int_{M_4} \mathrm{Tr}_{Sp(k)}F^2=2n_2+1
\end{equation}
where $S^1$ has periodic spin structure and $n_1, n_2\in \mathbb{Z}$. We have used a normalized trace $\frac{1}{4}\mathrm{Tr}_{G}F^2\equiv \frac{1}{4h^\vee_{G}}\mathrm{Tr}_{Adj.}F^2$ which canonically integrates to $1$ for a single instanton.

\begin{table}[t!]
\centering
\begin{tabular}{|c||c|c|c|c|c|c|c|c|}
\hline
 $i$ & $0$ & $1$ & $2$ & $3$ & $4$ & $5$ & $6$ & $7$  \\
 \hline\hline
 $KO^i(\mathrm{pt})$ & $\mathbb{Z}$ & $0$ & $0$  & $0$ & $\mathbb{Z}$ & $0$  & $\mathbb{Z}_2$ & $\mathbb{Z}_2$ \\
 \hline
\end{tabular}
\caption{$KO$-groups of a point which are periodic modulo $8$.}
\label{table:frozendessert}
\end{table}

Such a 9D TFT was explicitly given in \cite{Garcia-Etxebarria:2017crf} and can be constructed as follows. Given a 9-manifold $N_9$, consider an $\mathfrak{sp}(k)$ principle bundle over $N_9$, $P$, and its
associated vector bundle in the fundamental representation, $V$. This vector bundle defines a class in K-theory $\xi:=[V]\in KSp^0(N_9)=KO^{4}(N_9)$ where we have used Bott periodicity $KSp^i=KO^{i+4}=KSp^{i+8}=KO^{i-4}$. This class $\xi$ can be integrated over any dimension-$n$ submanifold of $N_9$ equipped with spin structure
\begin{equation}
    \int_{M_n\subset N_9} \xi\in KO^{4-n}(\mathrm{pt})
\end{equation}
where the groups $KO^{*}(\mathrm{pt})$ are given as in Table \ref{table:frozendessert}. The argument of the previous paragraph implies that the 9D SymTFT action is then
\begin{equation}\label{eq:symtfttocancel}
    S_{9D}=\int_{N_9} \xi^2\in KO^{-1}(\mathrm{pt})=\mathbb{Z}_2
\end{equation}
where $\xi^2:=\xi\cup \xi\in KO^0(N_9)$ is the cup product defined by the ring structure of $KO$. To understand this, consider the decomposition $N_9=M_5\times \Sigma_4$ and an $Sp(k)$ bundle $P$ such that there is a charge-1 instanton gauge configuration on $\Sigma_4$. In K-theory terms, this means that $\int_{\Sigma_4} \xi=1\in KO^0(\mathrm{pt})= \mathbb{Z}$. From the previous paragraph, we also have a non-trivial mod-2 index for the zero-mode Dirac operator on $M_5$ which in K-theory language can be expressed as $\int_{M_5} \xi=1\in KO^{-1}(\mathrm{pt})= \mathbb{Z}_2$. Therefore having $\int_{N_9}\xi^2=1 \; \mathrm{mod}\;  2$ in this background captures the 8D gauge anomaly.

\section{The Topological Green-Schwarz Mechanism}\label{sec:topgs}

From the anomalous backgrounds \eqref{eq:backgroundanom}, we see that coupling the 8D $\mathfrak{sp}(k)$ gauge theory to a TFT such that the sum over gauge configurations is restricted to even charged instantons, i.e. those with $\int \frac{1}{4}\mathrm{Tr}_{Sp(k)}F^2\in 2\mathbb{Z}$, would act as a topological Green-Schwarz mechanism. We will now show roughly how such a restriction takes place from topological terms present in the 9D SymTFT, before a more careful treatment in Section \ref{ssec:halfd3s}.

Since the minimal charge instanton on stack of $k$ D7-branes probing a $O7^+$-plane is a $\frac{1}{2}D3$-brane when $k>1$, this means that the topological coupling must restrict the D3-brane charge to be an integer along the orientifold plane. A restriction on the sum over instantons has appeared in field theory literature before \cite{Seiberg:2010qd,Tanizaki:2019rbk} so we can at first try to emulate the steps taken there. For a pure 4D YM theory with group $G$ such a coupling that restricts the instanton sum was given in \cite{Tanizaki:2019rbk} as
\begin{equation}\label{eq:instrestr}
  \int_{M_4}c_0\wedge\left(p da_3+\frac{1}{4}\mathrm{Tr}_{G}F^2\right)
\end{equation}
where $c_0$ is periodic scalar $c_0\sim c_0 +1$, $a_3$ is a $U(1)$ 3-form potential, and $p$ is some integer. Since $\int_{M_4}da_3\in \mathbb{Z}$ in our normalization, the equation of motion for $c_0$ restricts the instanton number to be a multiple of $p$
\begin{equation}\label{eq:instresr2}
  \frac{1}{4}\mathrm{Tr}_{G}F^2\in p\mathbb{Z},
\end{equation}
and the equation of motion for $a_3$ imposes $dc_0=0$ so $c_0$ is not a propagating degree of freedom (even if one includes a kinetic term for it). For the 8D $\mathfrak{sp}(k)$ gauge theory, one would thus expect the term
\begin{equation}\label{eq:intsrestr3}
   \int_{M_8}\frac{1}{2}c_4\wedge\left(2 da_3+ \frac{1}{4}\mathrm{Tr}_{Sp(k)}F^2\right)
\end{equation}
to restrict to a sum over an even charge instantons; we will motivate the factor of $1/2$ momentarily. Indeed, the second term resembles the coupling
\begin{equation}\label{eq:intsrestr4}
   \int_{M_8}C_4\wedge \left( \frac{1}{2}\times \frac{1}{4}\mathrm{Tr}_{Sp(k)}F^2\right)
\end{equation}
that is known to arise in the Wess-Zumino (WZ) action of the $O7^++kD7$-brane stack. Including the additional factor of $1/2$ in \eqref{eq:intsrestr3} is motivated by the $1/2$ in \eqref{eq:intsrestr4} which can be derived by the fact that as $k$ D7-branes approach a $O7^+$-plane, the gauge group enhances as $U(k)\hookrightarrow Sp(k)$ where $U(k)$ embeds as a Dynkin index-2 subgroup when $k>1$\cite{Atiyah:2001qf}\footnote{This reference argues for such a factor of $1/2$ for the analogous term on a $O6^+$/$D6$-brane stack.}. In other words, charge-2 instantons of $Sp(k)$ appear as charge-1 instantons in the canonical normalization of $U(k)$.

The key difference between \eqref{eq:intsrestr4} and the second term of \eqref{eq:intsrestr3} then is that $c_4$ is summed over in the path integral while $C_4$ is typically taken to be a fixed background field. We say typically because, while $C_4$ is certainly not associated to a gapless 4-form potential degree of freedom on the 7-brane stack, we are free to choose Neumann boundary conditions for $C_4$ along the asymptotic boundary of the spatial directions normal to the 7-brane stack, i.e. along $S_\infty^1=\partial \mathbb{C}_\perp$. This boundary condition can then reproduce the second term in \eqref{eq:intsrestr3} if such boundary conditions are chosen.

As for the $c_4\wedge da_3$ term, we first notice that the 9D SymTFT action contains the coupling\footnote{Similar terms have recently appeared in the SymTFT literature for describing SymTFTs for $U(1)$ symmetries \cite{Brennan:2024fgj,Antinucci:2024zjp}.}
\begin{equation}\label{eq:9dsymtft}
  \int_{N_9} F_5\wedge dA_3 \subset S_{9D}
\end{equation}
where $A_3$ is defined from reducing $C_4$ along $S^1_\infty$ (with coordinate\footnote{If one takes a metric normal to the 7-brane which takes into account the gravitational backreaction then there is a deficit angle for the natural angular coordinate but we will take the convention that $\phi$ is appropriately scaled to have $2\pi$ periodicity.} $\phi\sim \phi+2\pi$):
\begin{equation}
    C_4\supset A_3\wedge d\phi
\end{equation}
This term is simply the dimensional reduction of the ``kinetic term" for $F_5$ on $S^1_\infty$. We used scare quotes because the naive 10D kinetic term $F_5\wedge *F_5$ vanishes because $F_5$ is self-dual so one must treat this term with some care\footnote{One such approach is to extend the IIB to one higher-dimension \cite{Belov:2006xj}, see also the recent \cite{GarciaEtxebarria:2024fuk}}. However, because the dimensional reduction of this kinetic term on $S^1_\infty$ has no such difficulties we are just left with \eqref{eq:9dsymtft}. We can then simply inflow \eqref{eq:9dsymtft} to the 8D worldvolume of the $\mathfrak{sp}(k)$ gauge theory to $C_4\wedge dA_3$ which reproduces $c_4\wedge da_3$ if we also assume $A_3$ has Neumann boundary conditions.

In summary, we see then that we are left with an anomaly-free 8D gauge theory provided $C_4$ and $A_3$ have Neumann boundary conditions along $S^1_\infty$ which begs the question: what if one chooses Dirichlet boundary conditions for either of these fields? After all, that certain asymptotic boundary conditions are forced due to anomaly cancellation of a field theory would not be unique to this situation. For instance, in the case of 5D SCFTs constructed from M-theory on $\mathbb{C}^3/\mathbb{Z}_N$, one can have a cubic 1-form anomaly \cite{Apruzzi:2021nmk} (see also \cite{DelZotto:2024tae}). This forces a background 1-form field $B_2$ (which originates from the M-theory $C_3$) to have Dirichlet boundary conditions while its magnetic dual $B_3$ (which originates from $C_6$) is forced to have Neumann boundary conditions if one wants a well-defined partition function for the 5D SCFT. However, with a bit more care, we will be able to conclude that the 8D $\mathfrak{sp}(k)$ gauge anomaly cancels regardless of which boundary conditions one chooses for $C_4$ and $A_3$.

%In case people complain about this: a D3 off at infinity still creates an instanton in the 7-brane stack so it still sees C_4 at infinity whose boundary conditions we are free to pick.

\subsection{Symmetry Operators from (Flux)branes}\label{ssec:symopsflux}
To better understand the anomaly cancellation mechanism for arbitrary choices of boundary conditions for $C_4$ and its reduction along $S^1_\infty$, we focus on four topological operators present in the 9D SymTFT of the 8D gauge theory.

Following \cite{Heckman:2022muc} we can form symmetry operators for these theories by wrapping branes along cycles of the asymptotic boundary. For the case at hand, we will focus on wrapping $D3$-branes at infinity and there are only two cycles of $S_\infty^1$ to consider: a circle and a point, so we have the following topological operators
\begin{equation}
    D3(\mathrm{pt})\; \leftrightarrow \;  \mathcal{U}(M_4), \quad \quad \quad D3(S^1)\; \leftrightarrow \; \mathcal{V}(M_3).
\end{equation}
Here, $M_4$ and $M_3$ represent the 4 and 3-manfiolds which are the remaining directions of the worldvolume of the $D3$s after compactifying on $\mathrm{pt}$ or $S^1$ respectively. Including multiple D3s amounts to products of the topological operators\footnote{In this paper, we will always take fusion/product of topological operators as a field theorist would and do not consider the effects of non-abelian gauge enhancement of multiple D-branes at infinity coinciding. Physically, this just means that we take the D3 worldvolumes to coincide in the 8D gauge theory or 9D SymTFT directions, but not in the additional spatial directions. For more detailed remarks on this subtlety see \cite{Bah:2023ymy}.}, i.e. $n\times D3(\mathrm{pt})\; \leftrightarrow \;  \mathcal{U}^n(M_4)$. From the point of view of the 8D gauge theory, $\mathcal{U}(M_4)$ generates a 3-form symmetry and $\mathcal{V}(M_3)$ a 4-form symmetry and are a priori non-invertible. This is due to that presence of topological terms on the D3 worldvolume (which we assume to have trivial Pontryagin class) such as
\begin{equation}\label{eq:wzbrane}
    \int_{D3}\sum_{i\; \mathrm{even}}C_i \wedge \exp(F_2-B_2) \subset S^{D3}_{top.}
\end{equation}
which implies that $\mathcal{U}$ and $\mathcal{V}$ are each the product of several TFTs which will not be of importance for this work\footnote{The most of these factors are condensation defects which makes these operators ``weakly invertible" in the terminology of \cite{Heckman:2024obe}.}. In addition to the ``standard" Wess-Zumino terms of \eqref{eq:wzbrane}, there are additional terms that result from imposing that the D3 is invariant under S-duality as well as an additional term sensitive to the presence of a $\frac{1}{2} D3$ brane in an $O7^+$ plane as we will see later.

\begin{figure}
\centering
\includegraphics[scale=0.30, trim = {0cm 0cm 0cm 4cm}]{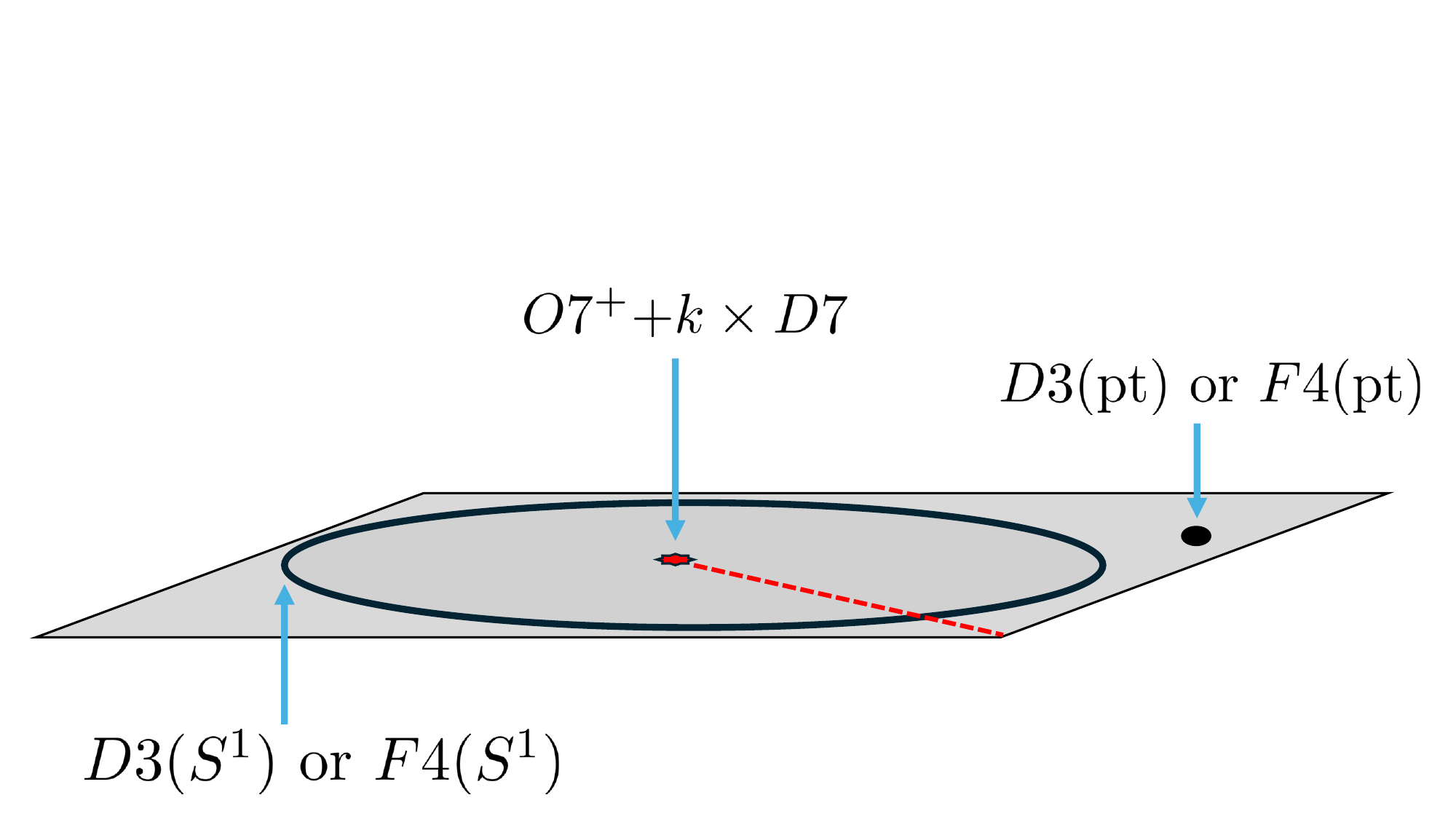}
\caption{Here we illustrate the transverse $\mathbb{C}_\perp$ directions to the $O7^+$+$k\times D7$ brane system. We engineer the topological operators of interest by wrapping a D3-brane or F4-brane on a point or 1-cycle of the asymptotic $S^1_\infty$. The red dotted line indicates a cut for the $SL(2,\mathbb{Z})$ monodromy of the 7-brane stack.}
\label{fig:topops}
\end{figure}

We can also obtain topological operators by wrapping a fluxbrane on $\mathrm{pt}$ or $S^1$ \cite{Cvetic:2023plv}. Of concern for us will be a flux 4-brane which we refer to as $F4$ which is the analog of a flux tube for the $F_5$ RR flux\footnote{Fluxbranes have been previously considered in the string theory literature \cite{Gutperle:2001mb,Emparan:2001gm} and one can write down non-supersymmetric supergravity solutions which are higher-dimensional/higher-form analogs of the Melvin universe solution in 4D \cite{Melvin:1963qx}.}. As such, the Wess-Zumino terms on a F4 fluxbrane are simply the exterior derivatives of the Wess-Zumino terms for the D3-brane, i.e. (for more details see \cite{Cvetic:2023plv})
\begin{equation}\label{eq:wzflux}
    \eta \int_{F4} d\left( \sum_{i\; \mathrm{even}}C_i \wedge \exp(F_2-B_2)\right)\subset S^{F4}_{top.}.
\end{equation}
The factor $\eta$ refers to the amount of flux localized on the $F4$. If we denote its worldvolume by $Y_5$ then the flux is given by
\begin{equation}
    *F_5=F_5=\eta \delta_{Y_5}.
\end{equation}
Note that this is an off-shell configuration of the RR-flux but since the $F4$ is infinitely far away from the 8D $\mathfrak{sp}(k)$ gauge theory so too is any information of its decay/dilution. We can then have the following topological operators in the 8D $\mathfrak{sp}(k)$ gauge theory and its 9D SymTFT (see also Figure \ref{fig:topops}):
\begin{equation}
    F4^\eta (\mathrm{pt})\; \leftrightarrow \; \mathcal{A}^\eta(M_5), \quad \quad \quad F4^{\eta}(S^1)\; \leftrightarrow \;  \mathcal{B}^\eta(M_4)
\end{equation}
Note that to properly reduce the terms \eqref{eq:wzbrane} and \eqref{eq:wzflux}, one has to use the fact that the $SL(2,\mathbb{Z})$ monodromy for an $O7^+$-plane is equivalent to that of an $O7^-$ plus 8 $D7$-branes (an $I^*_4$ singularity in F-theory language) which means that the monodromy matrix of the $O7^++k\times D7$ system is
\begin{equation}
    \begin{pmatrix}
        -1 & -(k+4)\\
        0 & -1
    \end{pmatrix}
\end{equation}
which gives a $C_0$ monodromy of $C_0(\phi+2\pi)= C_0(\phi)+(k+4)$. This implies that we have a background solution
\begin{equation}\label{eq:c0solution}
    C_0=\frac{(k+4)\phi}{2\pi}.
\end{equation}

In what follows, we will always assume that the background fields for the NS-NS fields and RR-fields other than $C_0$ (due to the 7-brane monodromy) and $C_4$ all vanish which greatly simplifies the actions of \eqref{eq:wzbrane} and \eqref{eq:wzflux}. Indeed, one can now present the topological operators from $D3$-branes as\footnote{Again, we note that we are ignoring the additional TFT factors that come from the additional terms in \eqref{eq:wzbrane}. For instance, the fact that $\int_{S^1\infty} dC_0=k+4$ implies that $\mathcal{V}(M_3)$ has a $U(1)$ Chern-Simons theory with level $-(k+4)$ localized on it.}
\begin{equation}\label{eq:d3topopsagain}
     \mathcal{U}^n(M_4)=\exp\left(2\pi in\int_{M_4}C_4\right), \quad \quad \mathcal{V}^m(M_3)=\exp\left(2\pi im\int_{M_3}A_3\right).
\end{equation}

Similarly, we can present the topological operators coming from the $F4$-branes as
\begin{equation}\label{eq:f4topopsagain}
     \mathcal{A}^{\eta}(M_5)=\exp\left(2\pi i\eta \int_{M_5}F_5\right), \quad \quad \mathcal{B}^{\eta'}(M_4)=\exp\left(2\pi i \eta' \int_{M_4}G_4\right)
\end{equation}
where $G_4:=dA_3$.

%\begin{equation*}
%    \mathcal{C}_{\mathcal{B}}= \sum^{\infty}_{\ell=-\infty} \exp\left(i\eta' (k+4) \ell\right)
%\end{equation*}

To summarize so far, we have operators $\mathcal{U}$, $\mathcal{V}$, $\mathcal{A}^\eta$, $\mathcal{B}^\eta$ which are essentially (save for the presence of the condensation defect factors which we ignore) symmetry operators for the higher-form symmetries $\mathbb{Z}^{(3)}$, $\mathbb{Z}^{(4)}$, $U(1)^{(2)}$, and $U(1)^{(3)}$ respectively. Since fluxbranes by construction measure the flux of $D3$ branes, $\mathcal{U}$ and $\mathcal{B}^\eta$ will have non-trivial commutation relations with each other and likewise for $\mathcal{V}$ and $\mathcal{A}^\eta$. More precisely we have the following braiding relations in 9D SymTFT (for more details see \cite{Cvetic:2023plv})
\begin{equation}\label{eq:link1}
   \mathcal{U}^{n}(M_4)\mathcal{B}^\eta(M'_4)=e^{2\pi i \eta n \mathrm{Link}(M_4,M'_4)}\mathcal{B}^\eta(M'_4)\mathcal{U}^{n}(M_4)
\end{equation}
and
\begin{equation}\label{eq:link2}
    \mathcal{V}^n(M_3)\mathcal{A}^\eta(M_5)=e^{2\pi i \eta n \mathrm{Link}(M_3,M_5)}\mathcal{A}^\eta(M_5)\mathcal{V}^n(M_3).
\end{equation}
Here $\mathrm{Link}(M_p,M_q)\in \mathbb{Z}$ ($p+q=8$) is the non-trivial Gauss link invariant for a 9-manifold. From the point of view of the 8D gauge theory, each commutation relation signifies a mixed 't Hooft anomaly: first one between $\mathbb{Z}^{(3)}$ and $U(1)^{(3)}$, and second between $\mathbb{Z}^{(4)}$ and $U(1)^{(2)}$.

We are now ready to properly address the possible boundary conditions of these operators and, in turn, for the RR-potentials/fluxes along the asymptotic boundary $S^1_\infty$. This asymptotic boundary is assigned a Hilbert space by IIB and the topological operators above act on it. The commutation relation \eqref{eq:link1}, for instance,  tells us that the relation between $\mathcal{U}^n$ and $\mathcal{B}^\eta$ eigenstates is equivalent to the problem of single particle quantum mechanics with a target $S^1$. Namely, if we take $X$ and $P$ to be the coordinate and momentum operators of such an $S^1$ then we roughly have:
\begin{equation}
     \mathcal{U}^n \; = \;  e^{2\pi in X}, \quad \quad \quad \mathcal{B}^\eta  \; = \; e^{2\pi i \eta P }.
\end{equation}
We recover the relations \eqref{eq:link1} if we have $\mathrm{Link}(M_4,M'_4)=1$.

More systematically, let us quantize the our 9D SymTFT on a manifold\footnote{We assume $H_4(X_8,\mathbb{Z})$ has no torsion for simplicity. } $X_8$ then we can take a basis of 4-cycles $M^{(i)}_4$, $i=1,...,b_4$, such that
\begin{equation}
    M^{(i)}_4 \cup M^{(j)}_4=Q_{ij}.
\end{equation}
where $Q_{ij}$ is the intersection matrix of $X_8$. Defining $\mathcal{U}^{n_i}_i=\mathcal{U}^{n_i}(M^{(i)}_4)$ and $\mathcal{B}^{\eta_i}_i=\mathcal{B}^{\eta_i}(M^{(i)}_4)$ we have that
\begin{equation}\label{eq:commrel}
    \mathcal{U}^n_i\mathcal{B}^\eta_j=\exp{\left(2\pi i \eta_i n_j Q_{ij}\right)} \; \mathcal{B}^\eta_j\mathcal{U}^n_i.
\end{equation}
Therefore if we denote $X_i$/$P_i$ as position/momentum operators for single particle quantum mechanics with target $T^{b_4}$, and define the vectors $\Vec{k}\in \mathbb{Z}^{b_4}$ and $\Vec{\eta}\in T^{b_4}$ then we can equate our topological operators with
\begin{equation}
    \prod^{b_4}_{i=1}\mathcal{U}_i^{k_i}=e^{2\pi i\Vec{k}\cdot \Vec{X}}, \quad \quad \quad \prod^{b_4}_{i=1} \mathcal{B}_i^{\eta_i}=e^{2\pi i\Vec{\eta}\cdot \Vec{P}}.
\end{equation}
As usual, we can form a position basis, $\ket{\Vec{x}}$, and a momentum basis $\ket{\Vec{p}}$ where
\begin{align*}
    e^{2\pi i\Vec{k}\cdot \Vec{X}}\ket{\Vec{x}}=e^{2\pi i\Vec{k}\cdot \Vec{x}}\ket{\Vec{x}}, \quad \quad \quad e^{2\pi i\Vec{\eta}\cdot \Vec{P}}\ket{\Vec{x}}=\ket{\Vec{x}+ Q\cdot \Vec{\eta}} \\
    e^{2\pi i\Vec{\eta}\cdot \Vec{P}}\ket{\Vec{p}}=e^{2\pi i\Vec{\eta}\cdot \Vec{p}}\ket{\Vec{p}}, \quad \quad \quad e^{2\pi i\Vec{k}\cdot \Vec{X}}\ket{\Vec{p}}=\ket{\Vec{p}-Q\cdot \Vec{k}}.
\end{align*}
Physically, the position basis signifies picking boundary conditions along $S^1_\infty$ such that we have definite values for the monodromies $\int_{M^{(i)}_4}C_4$ and the momentum basis signifies boundary conditions such that the fluxes $\int_{M^{(i)}_4}G_4=\int_{M^{(i)}_4\times S_\infty^1}F_5$ have definite values. Choosing a position basis is equivalent to gauging the $\mathbb{Z}^{(3)}$ symmetry generated by the $\mathcal{U}$ operators and choosing a momentum basis is equivalent to gauging the $U(1)^{(3)}$ symmetry generated by the $\mathcal{B}^\eta$ operators. This is most clear by considering that these bases are related by a Fourier transform
\begin{equation}\label{eq:fourier}
  \ket{\Vec{x}}=\frac{1}{(2\pi)^{b_4}}\sum_{\Vec{p}}e^{2\pi i \Vec{p}\cdot \Vec{x}}\ket{\Vec{p}}.
\end{equation}
In summary then
\begin{align*}\label{eq:summary}
  \ket{\Vec{x}} \; \textnormal{basis} \quad \Leftrightarrow \quad  \textnormal{Dirichlet b.c. (resp. Neumann) for $C_4$ ($G_4$)} \quad \Leftrightarrow \quad  U(1)^{(3)} \; \textnormal{symmetry}\\
  \ket{\Vec{p}} \; \textnormal{basis} \quad \Leftrightarrow \quad \textnormal{Dirichlet b.c. (resp. Neumann) for $G_4$ ($C_4$)}  \quad \Leftrightarrow \quad  \mathbb{Z}^{(3)} \; \textnormal{symmetry}
\end{align*}
where we have also indicated which global symmetry survives (i.e. is not gauged) for a given boundary condition. Together choosing various (possibly mixed between the different directions of $T^{b_4}$) position/momentum bases amounts to picking the boundary state $\ket{\mathcal{B}_{\mathrm{top}}}$ which will be used in defining the 8D partition function as outlined in the Introduction. Also, note that the braiding \eqref{eq:link2} also means we can pick an independent position and momentum basis for $\mathcal{V}^n$ and $\mathcal{A}^\eta$ but these will not be needed in our main argument. 

Finally, we mention these are quite similar to the steps taken in recent works on characterizing $U(1)$ symmetries in the context of the SymTFT paradigm \cite{Brennan:2024fgj,Antinucci:2024zjp,DelZotto:2024tae,Apruzzi:2024htg} and it would be interesting to sharpen the correspondence with these works.

\subsection{$\frac{1}{2}D3$-branes, a Phase Ambiguity, and Anomaly Cancellation}\label{ssec:halfd3s}
Before deriving the condition that the 8D $\mathcal{N}=1$ $\mathfrak{sp}(k)$ has only even-charged instantons and that the anomaly \eqref{eq:symtfttocancel} is canceled, we must first understand how the operators $\mathcal{U}$ and $\mathcal{B}^\eta$ are affected by the presence of a $\frac{1}{2}D3$ which sources a unit charge $\mathfrak{sp}(k)$ instanton. In such a scenario, we have open strings stretching between the $\frac{1}{2}D3$ and the $D3$- or $F4$-brane which transform in a bifundamental representation of $\mathfrak{sp}(k)\times \mathfrak{u}(1)$. That open strings can end on fluxbranes is clear from Appendix A of \cite{Cvetic:2023plv} where it was shown that fluxbranes can be realized as non-trivial configurations of $D$-brane/anti-$D$-brane pairs of higher dimension. From the point of view of the $D3$ or $F4$ off at asymptotic infinity this string mode, while infinitely massive, transforms under an $\mathfrak{sp}(k)$ flavor vector bundle over the $D3$ or $F4$ worldvolume. In particular, this flavor bundle is sensitive to the presence of (fractional) $D3$-branes on the $O7^++kD7$-brane system which are transverse to the $D3$ or $F4$ worldvolume. More specifically if we take a D3 at infinity, $\mathcal{U}(M_4)$, and a $\frac{1}{2}D3$ supported on a submanifold $\Sigma_4$ of the 7-brane stack such that $\Sigma_4$ and $M_4$ intersect at a single point\footnote{More generally, we are interested in cases where $\Sigma_4\cup M_4$ is an odd number and the same conclusions below will follow. If this intersection number is even then a $\frac{1}{2}D3$-brane on $\Sigma_4$ induces an even charged instanton background on $M_4$ which is non-anomalous and so does not concern us. We thank M. Dierigl for asking about this subtlety.}, $\Sigma_4\cup M_4=1$. Then $\mathcal{U}(M_4)$ has a topological  $\mathfrak{sp}(k)$ vector bundle with a unit instanton localized on it $\int_{M_4}\frac{1}{4}\mathrm{Tr}_{Sp(k)}F^2=1$. A similar remark also applies to the symmetry operator arising from the F4-brane on $S^1_\infty$, $\mathcal{B}^\eta(M_4)$. 

\begin{figure}
\centering
\includegraphics[scale=0.30, trim = {0cm 0cm 0cm 0cm}]{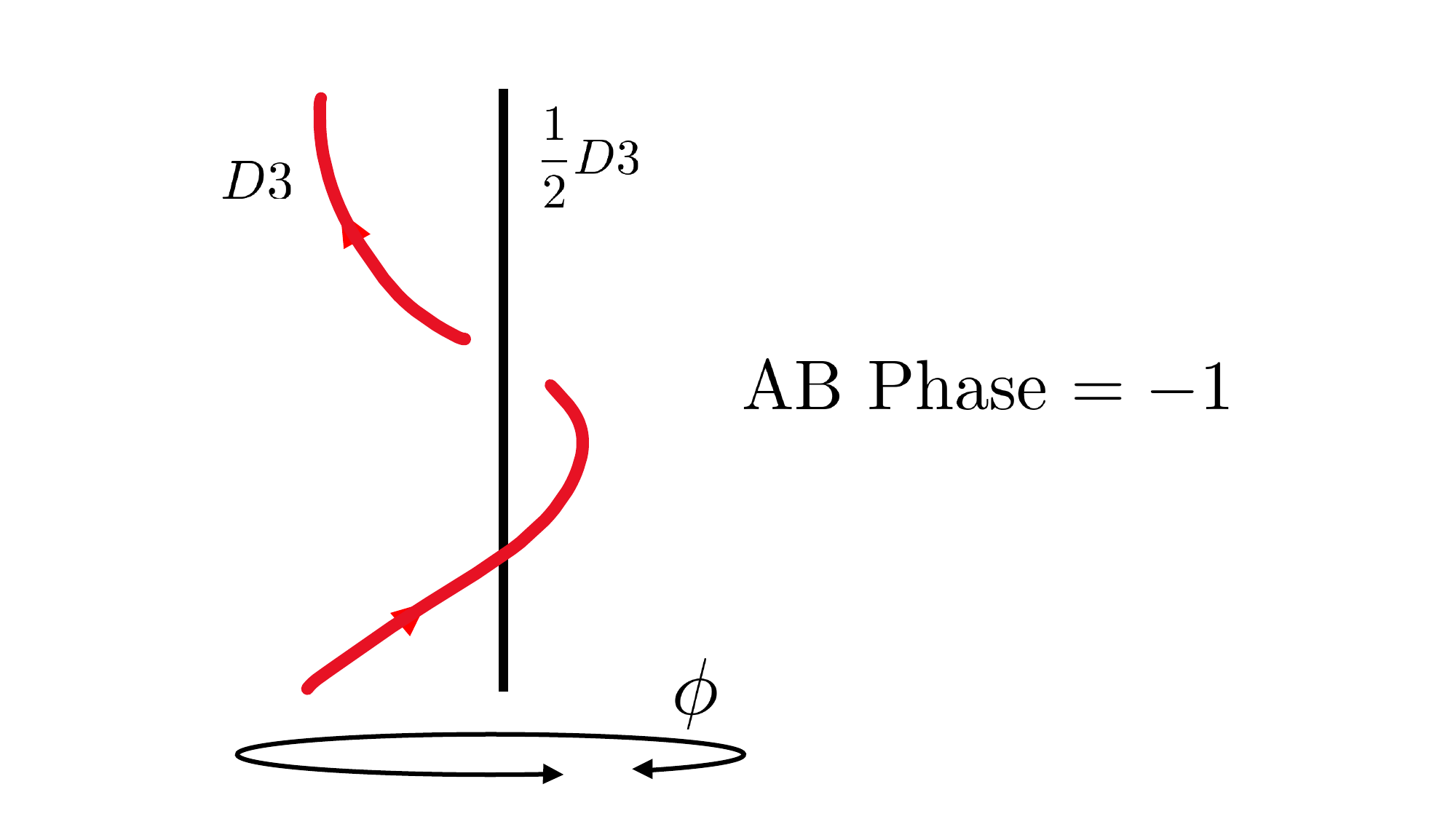}
\caption{Braiding between $D3$ and $\frac{1}{2}D3$ generating a non-trivial Aharanov-Bohm phase. The latter brane is fixed to the $O7^+$-plane locus (not pictured).}
\label{fig:D3braiding}
\end{figure}

Given then that $\mathcal{U}$ and $\mathcal{B}^\eta$ are each equipped with a $\mathfrak{sp}(k)$ vector bundle, we now wish to understand how they couple to it. Indeed for $\mathcal{U}$ this is straightforward to answer as it is induced from the 7-brane coupling
\begin{equation}\label{eq:trf22}
    \propto\frac{1}{4^2}\int_{X_8}C_0 \mathrm{Tr}_{Sp(k)}F^2\wedge \mathrm{Tr}_{Sp(k)}F^2
\end{equation}
because, after all, $\mathcal{U}$ is engineered from a D3 which is a charge-2 instanton in the $\mathfrak{sp}(k)$ gauge theory. Taking it to be infinitely far away from the 7-brane amounts to taking a charge-2 instanton infinitely far away from the origin of its Coulomb branch. This means that we have a worldvolume coupling $\propto \frac{1}{2}\int_{M_4}C_0 \mathrm{Tr}_{Sp(k)}F^2$ in the $\mathcal{U}(M_4)$ action. We can solve for the proportionality factor of \eqref{eq:trf22} by considering that in the presence of $\frac{1}{2}D3(\Sigma_4)$, $\mathcal{U}(M_4)$ has an Aharonov-Bohm (AB) phase as we move it around $S^1_\infty$, see Figure \ref{fig:D3braiding}. This phase is given by the exponential of the Dirac pairing between the $D3$ and $\frac{1}{2}D3$, $\exp(2\pi i (1)\times \frac{1}{2})=-1$. Using the solution for $C_0$ in the 7-brane background, \eqref{eq:c0solution}, we see that the $\mathfrak{sp}(k)$ instanton background modifies $\mathcal{U}(M_4)$ as
\begin{equation}\label{eq:mod1}
  \mathcal{U}(M_4)=\exp\left(2\pi i\int_{M_4}C_4\right)\cdot \exp\left(\frac{2\pi i}{2(k+4)}\times \frac{1}{4}\int_{M_4}C_0 \mathrm{Tr}_{Sp(k)}F^2\right).
\end{equation}
Similarly, $\mathcal{B}^{\eta}(M_4)$ is now
\begin{equation}\label{eq:mod2}
    \mathcal{B}^{\eta}(M_4)=\exp\left(2\pi i \eta \int_{M_4}G_4\right)\cdot \exp\left(\pi i \eta \times \frac{1}{4}\int_{M_4} \mathrm{Tr}_{Sp(k)}F^2\right).
\end{equation}

We see that the periodicity of $\eta$ in $\mathcal{B}^{\eta}(M_4)$ is extended to $\eta\sim \eta+2$ when there are an odd number of instantons and if we let $N:=\frac{1}{4}\int_{M_4} \mathrm{Tr}_{Sp(k)}F^2\in \mathbb{Z}$ then we can identify
\begin{equation}\label{eq:mod3}
    \mathcal{B}^{\eta=1}(M_4)=(-1)^N
\end{equation}
as the instanton parity operator which is associated with a 4-cycle $M_4$ whose dependence we suppress in the notation. If we now label $\mathcal{U}$ by its position $\phi\in S^1_\infty$, then it follows from \eqref{eq:c0solution} that the shift operation $P_{2\pi}:\phi\rightarrow \phi+2\pi$ transforms $\mathcal{U}^{(\phi)}$ as
\begin{equation}\label{eq:rotationbraiding}
    P_{2\pi}\mathcal{U}^{(\phi)}P_{2\pi}=(-1)^N\mathcal{U}^{(\phi)}.
\end{equation}
due to the non-trivial linking around $\frac{1}{2}D3$-branes. Such an operation might seem obscure from the perspective of the 8D gauge theory given that it involves the transverse dimensions, but we claim that it is simply fermion parity
\begin{equation}\label{eq:FeqN}
    P_{2\pi}=(-1)^F.
\end{equation}
This simply follows from the decomposition of the 10D IIB supercharges and the fact that a $2\pi$ rotation in any 2-plane is equivalent to the action of fermion parity. More specifically, the 10D IIB supercharges can be arranged into a complex 10D Weyl fermion which decomposes as
\begin{align}
    \mathfrak{spin}(9,1)\supset \mathfrak{spin}(7,1)\times \mathfrak{spin}(2)\\
    \mathbf{16}^L\quad \longrightarrow \quad  \mathbf{8}^L_{+1/2}\oplus \mathbf{8}^R_{-1/2}. \quad  \quad
\end{align}
The 7-brane system only preserves the 8D left-handed Weyl fermion $\mathbf{8}^L_{+1/2}$. The $\mathfrak{spin}(2)\simeq \mathfrak{u}(1)$ factor is the R-symmetry for the 8D $\mathcal{N}=1$ gauge theory, so the $+1/2$ charge means that a rotation in the transverse plane $\phi\rightarrow \phi+2\pi$ acts by $(-1)$ on the gauge theory supercharge and the 8D adjoint Majorana fermion\footnote{We are also implicitly relying on the fact that the $SL(2,\mathbb{Z})$ monodromy matrix does not produce an additional action on the IIB supercharges which is possible if the bottom left entry in the matrix is non-trivial, see \cite{Debray:2021vob} for details on this point.}. Similar to 4D Minkowski space fermions, this can equivalently be packaged as an 8D Weyl fermion which makes a $U(1)$ R-symmetry action more manifest. The braiding \eqref{eq:rotationbraiding} then tells us that unless $(-1)^N=1$, there is a mixed anomaly between fermion parity and $\mathbb{Z}^{(3)}$:
\begin{equation}\label{eq:ufmixedanom}
    (-1)^F\mathcal{U}=(-1)^N\mathcal{U} (-1)^F.
\end{equation}
Such a relation can be modified if we take the geometry normal to the 7-brane to be the backreacted geometry rather than $\mathbb{C}$, we come back to this point later in the section.

We now return to the construction of position/momentum boundary states to see how the presence of $\frac{1}{2}D3$-branes affects the conclusions of Section \ref{ssec:symopsflux}. We will address the case of \textit{fixed} $\phi$ first. From \eqref{eq:commrel} we see that $(-1)^N$ has a trivial commutation relation with $\mathcal{U}^{(\phi)}$ because its braiding phase with $\mathcal{B}^\eta$ is invariant under $\eta\rightarrow \eta +1$. Since $\mathcal{B}^\eta$ also has a trivial commutation relation with $(-1)^N$, one can conclude that we can form position/momentum eigenstates that are also eigenstates under instanton parity. We denote these states by $\ket{x,N}:=\ket{x}\otimes \ket{N}$ and $\ket{p,N}:=\ket{p}\otimes \ket{N}$ where $\ket{N}$ is defined to be a state
with a charge-$N$ $\mathfrak{sp}(k)$ instanton configuration along $\Sigma_4$, and if we fix $\phi=\phi_0$, then $\mathcal{B}^\eta$ and $\mathcal{U}^{(\phi_0)}$ act as
\begin{align}
    \mathcal{B}^{\eta}\ket{x,N}=e^{i\pi \eta N}\ket{x+ \eta ,N}, \quad \quad \quad \mathcal{B}^{\eta}\ket{p,N}=e^{2\pi i \eta p}e^{i \pi \eta N}\ket{p,N}\\
    \mathcal{U}^{(\phi_0)}\ket{x,N}=e^{2\pi i x}e^{i\phi_0N/2}\ket{x,N}      \quad \quad \quad   \mathcal{U}^{(\phi_0)}\ket{\Vec{p}, N}=e^{i\phi_0N/2}\ket{p-1, N}.
\end{align}

We can now vary $\phi$ to see how the above states are acted upon by the shift $P_{2\pi}=(-1)^F$. We can first rewrite this action on the position states as
\begin{equation}\label{eq:posNaction}
  (-1)^F\ket{x,N}=(-1)^F\mathcal{U}^{(\phi)}e^{-2\pi i x}e^{-i\phi_0N/2}\ket{x,N}=\mathcal{U}^{(\phi)}e^{-2\pi i x}e^{-i\phi_0N/2}(-1)^N(-1)^F \ket{x,N}
\end{equation}
and similarly for the momentum states
\begin{equation}\label{eq:momNaction}
  (-1)^F\ket{p,N}=(-1)^F\mathcal{U}^{(\phi)}e^{-i\phi N/2}\ket{p+1,N}=\mathcal{U}^{(\phi)}e^{-i\phi N/2}(-1)^N(-1)^F\ket{p+1,N}
\end{equation}
where we have used the fact that $(-1)^F$ and $(-1)^N$ commute which is clear from \eqref{eq:mod2}. Importantly, we must now consider whether $\ket{x,N}$ and/or $\ket{p,N}$ are eigenstates under $(-1)^F$. If they are not, then the boundary states will not preserve supersymmetry because supercharges anticommute with $(-1)^F$, states in a representation of this algebra which are not eigenstates of $(-1)^F$ will have non-trivial vacuum expectation values for supercharges. When $\ket{\mathcal{B}_\mathrm{top}}$ does not preserve supersymmetry then neither will the 8D theory $\mathcal{T}_{\mathfrak{sp}(k)}\otimes \mathrm{TFT}^{(\mathcal{B}_{\mathrm{top}})}_{8D}$.

Given then that we must assume that $\ket{x,N}$ and $\ket{p,N}$ are eigenstates under $(-1)^F$, let us now focus for concreteness on a $(-1)^F=\pm 1$ position states: $\ket{x,N,\pm}$. In these cases we have
\begin{equation*}\label{eq:posNaction2}
 \ket{x,N,+}=(-1)^F\ket{x,N,+}=\mathcal{U}^{(\phi)}e^{-2\pi i x}e^{-i\phi_0N/2}(-1)^N(-1)^F \ket{x,N,+}=(-1)^N\ket{x,N,+}
\end{equation*}
and
\begin{equation*}\label{eq:posNaction3}
 -\ket{x,N,-}=(-1)^F\ket{x,N,-}=\mathcal{U}^{(\phi)}e^{-2\pi i x}e^{-i\phi_0N/2}(-1)^N(-1)^F \ket{x,N,-}= -(-1)^N\ket{x,N,-}
\end{equation*}
respectively. We learn then that these supersymmetric states must obey the condition
\begin{equation}\label{eq:keycond}
  (-1)^N=+1
\end{equation}
as they would be projected out by above manipulations otherwise. Similar remarks equally apply to the momentum states using \eqref{eq:momNaction} and as well as to boundary states which do not preserve supersymmetry but still conserve $(-1)^F$\footnote{These would transform under $(-1)^F$ as $(-1)^F\ket{a}=\ket{b}$, $(-1)^F\ket{b}=\ket{a}$ where $a$ and $b$ are labels independent of position, momentum, and instanton number. Boundary states that do not conserve $(-1)^F$ are disallowed by the fact that the theory is fermionic.}.

We therefore see that the 9D SymTFT automatically disallows configurations with odd $N$. This means that the 8D gauge anomaly is trivialized by the fact that the $\mathfrak{sp}(k)$ gauge configurations of the form \eqref{eq:backgroundanom} do not appear. Conceptually, what we have shown is that the mixed anomalies \eqref{eq:commrel} and \eqref{eq:ufmixedanom} require that any boundary state $\ket{\mathcal{B}_{\mathrm{top}}}$ we choose in order to have a well-defined 8D partition function satisfies $(-1)^N\ket{\mathcal{B}_{\mathrm{top}}}=+\ket{\mathcal{B}_{\mathrm{top}}}$. In terms of the K-theory discussion in Section \ref{sec:rev}, we can roughly think of this anomaly cancellation as the result of extending the 8D topological term \eqref{eq:trf22} to 9D and canceling the 9D anomaly theory action $\int_{N_9}\xi^2$. This is may be seen by the fact that $\int_{\Sigma_4}\xi=\int_{\Sigma_4}\frac{1}{4}\mathrm{Tr}_{Sp(k)}F^2\in \mathbb{Z}$ and that the K-theory class $\xi^2$ can be sensibly integrated over both 8- and 9-manifolds equipped with spin structure. It would be interesting to develop this perspective in future work.

\paragraph{Subtleties with 7-Brane Backreaction}
Up until this point we have been only focusing on the 7-brane stack without gravitational backreaction, so strictly speaking our analysis is rigorous when the string coupling vanishes, $g_s=0$. We claim that this is simply an artifact of working in the non-gravitational limit of the 7-brane gauge theory in the first place. Indeed since the
$\mathfrak{sp}(k)$ gauge coupling is $g^2_{SYM}\sim g_s(\alpha')^{-2}$, we see that the gauge theory is only fully decoupled from the gravitational degrees of freedom when $g_s=0$. 

In any case, if we consider our SymTFT construction for a 7-brane with the gravitational backreaction, i.e. a non-trivial deficit angle, then we can still see that consistent boundary conditions must satisfy $(-1)^N=+1$. Take for instance the minimally anomalous case of $\mathfrak{sp}(2)$. This is engineered from two $D7$-branes probing an $O7^+$ plane which has a deficit angle of $2 \pi$. Such a large deficit angle means that the 7-brane backreacts so much that the asymptotic circle $S^1_\infty$ becomes finite size. From the point-of-view of the elliptic F-theory geometry this is a $dP_9$ surface. The spin structure around $S^1_\infty$ is now \textit{periodic} which alters \eqref{eq:ufmixedanom} to
\begin{equation}
    \mathcal{U}=(-1)^N\mathcal{U}
\end{equation}
so there is a ``mixed anomaly" between $\mathcal{U}$ and the identity operator if $(-1)^N=-1$. Using similar arguments to before, position and momentum states with $(-1)^N=-1$ are projected out so we recover \eqref{eq:keycond}. For the case of more than two $D7$s probing an $O7^+$-plane, these can be Higgsed to the $\mathfrak{sp}(2)$ theory which implies they also are gauge anomaly-free.

\section{Conclusions}\label{sec:conc}

In understanding how the 8D $\mathcal{N}=1$ $\mathfrak{sp}(k)$ gauge theory is anomaly-free, we have seen the utility of both engineering topological operators from branes as well as the concept of topological symmetry operators with vector bundles. We expect that these tools can generally be useful for understanding subtle string theory anomaly cancellation conditions and global symmetry structures for brane/orientifold systems in lower dimensions as well. Similarly, these backgrounds are related to geometric singularities with fluxes localized on them which, depending on the context, are referred to as frozen singularities or discrete torsion orbifolds. For example, compactifying an $O7^+$-plane with $D7$ probes on a circle is equivalent to M-theory on a non-compact elliptic K3 with two frozen D-type singularities \cite{deBoer:2001wca,Tachikawa:2015wka} which each have localized fractional $G_4$ fluxes. Even for the ``basic" case of frozen ADE singularities in M-theory, there are still several basic properties of anomaly constraints and higher-form symmetries that are unknown, some of which are addressed in the upcoming \cite{FrozenToAppear}. Extracting such subtle data is generally outside of the scope of string perturbation theory which is why these topological methods can be quite useful for clarifying the non-perturbative behavior of string/M-/F-theory including perhaps scenarios with no supersymmetry, see for instance the recent \cite{Braeger:2024jcj}.

Additionally, in constructing the 9D SymTFT and 8D topological degrees of freedom of the 7-branes systems we have focused on, we have been far from exhaustive in describing the most general fusion algebras and categorical structure of operators, especially in the presence of background NSNS or RR fields. Similarly, because the topological operators we have discussed have vector bundles on them, they define Baum-Douglas classes of $KSp$-homology and we expect that one may have higher-fusion products from couplings such as $\int_{X_8}\xi_1\xi_2\xi_3$ where $\xi_i\in KSp^i(X_8)$ are the Poincar\'e duals of such classes. Also, we have not had anything to say on the $O7^+$-plane with no other 7-branes probing it but we expect it to be interesting 8D TFT in its own right. 

Finally, it would interesting to understand the interplay of these considerations when embedding in compact models whereby the possible Dirichlet/Neumann boundary conditions for $C_4$ and $G_4$ correlate the such choices on other 7-brane stacks. This is similar to how simple gauge groups can be mixed by a discrete quotient factors; for such considerations relevant to global structures of supergravity gauge groups see \cite{Cvetic:2023pgm,Gould:2023wgl,Cvetic:2022uuu}.

\section*{Acknowledgments}
We thank M. Dierigl, L. Lin, and H.Y. Zhang for helpful discussions and the Harvard Swampland Initiative for their hospitality during a visit in which a preliminary version of these results were presented. We also thank M. Dierigl for helpful comments on an earlier draft. We thank the SISSA Theoretical Particle Physics group for their hospitality during the completion of this work. We would like to especially thank C. Carter for encouragement and for suggesting the title of this paper. The work of ET is supported in part by the ERC Starting Grant QGuide-101042568 - StG 2021.

%\appendix

%\section{Instanton Restriction and Topological Operators}

\newpage

\bibliographystyle{utphys}
\bibliography{GSfrozen}

\providecommand{\href}[2]{#2}\begingroup\raggedright\begin{thebibliography}{10}

\bibitem{Garcia-Etxebarria:2017crf}
I.~n. Garc\'\i{}a-Etxebarria, H.~Hayashi, K.~Ohmori, Y.~Tachikawa, and
  K.~Yonekura, ``{8d gauge anomalies and the topological Green-Schwarz
  mechanism},'' \href{http://dx.doi.org/10.1007/JHEP11(2017)177}{{\em JHEP}
  {\bfseries 11} (2017) 177}, \href{http://arxiv.org/abs/1710.04218}{{\ttfamily
  arXiv:1710.04218 [hep-th]}}.

\bibitem{Cordova:2022ruw}
C.~Cordova, T.~T. Dumitrescu, K.~Intriligator, and S.-H. Shao, ``{Snowmass
  White Paper: Generalized Symmetries in Quantum Field Theory and Beyond},'' in
  {\em {Snowmass 2021}}.
\newblock 5, 2022.
\newblock \href{http://arxiv.org/abs/2205.09545}{{\ttfamily arXiv:2205.09545
  [hep-th]}}.

\bibitem{Schafer-Nameki:2023jdn}
S.~Schafer-Nameki, ``{ICTP Lectures on (Non-)Invertible Generalized
  Symmetries},'' \href{http://arxiv.org/abs/2305.18296}{{\ttfamily
  arXiv:2305.18296 [hep-th]}}.

\bibitem{Bhardwaj:2023kri}
L.~Bhardwaj, L.~E. Bottini, L.~Fraser-Taliente, L.~Gladden, D.~S.~W. Gould,
  A.~Platschorre, and H.~Tillim, ``{Lectures on generalized symmetries},''
  \href{http://dx.doi.org/10.1016/j.physrep.2023.11.002}{{\em Phys. Rept.}
  {\bfseries 1051} (2024) 1--87},
  \href{http://arxiv.org/abs/2307.07547}{{\ttfamily arXiv:2307.07547
  [hep-th]}}.

\bibitem{Luo:2023ive}
R.~Luo, Q.-R. Wang, and Y.-N. Wang, ``{Lecture Notes on Generalized Symmetries
  and Applications},''
\newblock 7, 2023.
\newblock \href{http://arxiv.org/abs/2307.09215}{{\ttfamily arXiv:2307.09215
  [hep-th]}}.

\bibitem{Shao:2023gho}
S.-H. Shao, ``{What's Done Cannot Be Undone: TASI Lectures on Non-Invertible
  Symmetry},'' \href{http://arxiv.org/abs/2308.00747}{{\ttfamily
  arXiv:2308.00747 [hep-th]}}.

\bibitem{Brennan:2024fgj}
T.~D. Brennan and Z.~Sun, ``{A SymTFT for Continuous Symmetries},''
  \href{http://arxiv.org/abs/2401.06128}{{\ttfamily arXiv:2401.06128
  [hep-th]}}.

\bibitem{Antinucci:2024zjp}
A.~Antinucci and F.~Benini, ``{Anomalies and gauging of U(1) symmetries},''
  \href{http://arxiv.org/abs/2401.10165}{{\ttfamily arXiv:2401.10165
  [hep-th]}}.

\bibitem{Bonetti:2024cjk}
F.~Bonetti, M.~Del~Zotto, and R.~Minasian, ``{SymTFTs for Continuous
  non-Abelian Symmetries},'' \href{http://arxiv.org/abs/2402.12347}{{\ttfamily
  arXiv:2402.12347 [hep-th]}}.

\bibitem{Apruzzi:2024htg}
F.~Apruzzi, F.~Bedogna, and N.~Dondi, ``{SymTh for non-finite symmetries},''
  \href{http://arxiv.org/abs/2402.14813}{{\ttfamily arXiv:2402.14813
  [hep-th]}}.

\bibitem{Apruzzi:2021nmk}
F.~Apruzzi, F.~Bonetti, I.~n.~G. Etxebarria, S.~S. Hosseini, and
  S.~Schafer-Nameki, ``{Symmetry TFTs from String Theory},''
  \href{http://arxiv.org/abs/2112.02092}{{\ttfamily arXiv:2112.02092
  [hep-th]}}.

\bibitem{Heckman:2022muc}
J.~J. Heckman, M.~H\"ubner, E.~Torres, and H.~Y. Zhang, ``{The Branes Behind
  Generalized Symmetry Operators},''
  \href{http://arxiv.org/abs/2209.03343}{{\ttfamily arXiv:2209.03343
  [hep-th]}}.

\bibitem{Heckman:2022xgu}
J.~J. Heckman, M.~Hubner, E.~Torres, X.~Yu, and H.~Y. Zhang, ``{Top down
  approach to topological duality defects},''
  \href{http://dx.doi.org/10.1103/PhysRevD.108.046015}{{\em Phys. Rev. D}
  {\bfseries 108} no.~4, (2023) 046015},
  \href{http://arxiv.org/abs/2212.09743}{{\ttfamily arXiv:2212.09743
  [hep-th]}}.

\bibitem{Zhang:2024oas}
H.~Y. Zhang, ``{K-theoretic Global Symmetry in String-constructed QFT and
  T-duality},'' \href{http://arxiv.org/abs/2404.16097}{{\ttfamily
  arXiv:2404.16097 [hep-th]}}.

\bibitem{Green:1984sg}
M.~B. Green and J.~H. Schwarz, ``{Anomaly Cancellation in Supersymmetric D=10
  Gauge Theory and Superstring Theory},''
  \href{http://dx.doi.org/10.1016/0370-2693(84)91565-X}{{\em Phys. Lett. B}
  {\bfseries 149} (1984) 117--122}.

\bibitem{Seiberg:2010qd}
N.~Seiberg, ``{Modifying the Sum Over Topological Sectors and Constraints on
  Supergravity},'' \href{http://dx.doi.org/10.1007/JHEP07(2010)070}{{\em JHEP}
  {\bfseries 07} (2010) 070}, \href{http://arxiv.org/abs/1005.0002}{{\ttfamily
  arXiv:1005.0002 [hep-th]}}.

\bibitem{Tanizaki:2019rbk}
Y.~Tanizaki and M.~\"Unsal, ``{Modified instanton sum in QCD and
  higher-groups},'' \href{http://dx.doi.org/10.1007/JHEP03(2020)123}{{\em JHEP}
  {\bfseries 03} (2020) 123}, \href{http://arxiv.org/abs/1912.01033}{{\ttfamily
  arXiv:1912.01033 [hep-th]}}.

\bibitem{Witten:1998wy}
E.~Witten, ``{AdS / CFT correspondence and topological field theory},''
  \href{http://dx.doi.org/10.1088/1126-6708/1998/12/012}{{\em JHEP} {\bfseries
  12} (1998) 012}, \href{http://arxiv.org/abs/hep-th/9812012}{{\ttfamily
  arXiv:hep-th/9812012}}.

\bibitem{Witten:1982fp}
E.~Witten, ``{An SU(2) Anomaly},''
  \href{http://dx.doi.org/10.1016/0370-2693(82)90728-6}{{\em Phys. Lett. B}
  {\bfseries 117} (1982) 324--328}.

\bibitem{Wang:2018qoy}
J.~Wang, X.-G. Wen, and E.~Witten, ``{A New SU(2) Anomaly},''
  \href{http://dx.doi.org/10.1063/1.5082852}{{\em J. Math. Phys.} {\bfseries
  60} no.~5, (2019) 052301}, \href{http://arxiv.org/abs/1810.00844}{{\ttfamily
  arXiv:1810.00844 [hep-th]}}.

\bibitem{Atiyah:2001qf}
M.~Atiyah and E.~Witten, ``{M theory dynamics on a manifold of G(2)
  holonomy},'' \href{http://dx.doi.org/10.4310/ATMP.2002.v6.n1.a1}{{\em Adv.
  Theor. Math. Phys.} {\bfseries 6} (2003) 1--106},
  \href{http://arxiv.org/abs/hep-th/0107177}{{\ttfamily arXiv:hep-th/0107177}}.

\bibitem{Belov:2006xj}
D.~M. Belov and G.~W. Moore, ``{Type II Actions from 11-Dimensional
  Chern-Simons Theories},''
  \href{http://arxiv.org/abs/hep-th/0611020}{{\ttfamily arXiv:hep-th/0611020}}.

\bibitem{GarciaEtxebarria:2024fuk}
I.~n. Garc\'\i{}a~Etxebarria and S.~S. Hosseini, ``{Some aspects of symmetry
  descent},'' \href{http://arxiv.org/abs/2404.16028}{{\ttfamily
  arXiv:2404.16028 [hep-th]}}.

\bibitem{DelZotto:2024tae}
M.~Del~Zotto, S.~N. Meynet, and R.~Moscrop, ``{Remarks on Geometric
  Engineering, Symmetry TFTs and Anomalies},''
  \href{http://arxiv.org/abs/2402.18646}{{\ttfamily arXiv:2402.18646
  [hep-th]}}.

\bibitem{Bah:2023ymy}
I.~Bah, E.~Leung, and T.~Waddleton, ``{Non-invertible symmetries, brane
  dynamics, and tachyon condensation},''
  \href{http://dx.doi.org/10.1007/JHEP01(2024)117}{{\em JHEP} {\bfseries 01}
  (2024) 117}, \href{http://arxiv.org/abs/2306.15783}{{\ttfamily
  arXiv:2306.15783 [hep-th]}}.

\bibitem{Heckman:2024obe}
J.~J. Heckman, J.~McNamara, M.~Montero, A.~Sharon, C.~Vafa, and I.~Valenzuela,
  ``{On the Fate of Stringy Non-Invertible Symmetries},''
  \href{http://arxiv.org/abs/2402.00118}{{\ttfamily arXiv:2402.00118
  [hep-th]}}.

\bibitem{Cvetic:2023plv}
M.~Cveti\v{c}, J.~J. Heckman, M.~H\"ubner, and E.~Torres, ``{Fluxbranes,
  generalized symmetries, and Verlinde\textquoteright{}s metastable
  monopole},'' \href{http://dx.doi.org/10.1103/PhysRevD.109.046007}{{\em Phys.
  Rev. D} {\bfseries 109} no.~4, (2024) 046007},
  \href{http://arxiv.org/abs/2305.09665}{{\ttfamily arXiv:2305.09665
  [hep-th]}}.

\bibitem{Gutperle:2001mb}
M.~Gutperle and A.~Strominger, ``{Fluxbranes in string theory},''
  \href{http://dx.doi.org/10.1088/1126-6708/2001/06/035}{{\em JHEP} {\bfseries
  06} (2001) 035}, \href{http://arxiv.org/abs/hep-th/0104136}{{\ttfamily
  arXiv:hep-th/0104136}}.

\bibitem{Emparan:2001gm}
R.~Emparan and M.~Gutperle, ``{From p-branes to fluxbranes and back},''
  \href{http://dx.doi.org/10.1088/1126-6708/2001/12/023}{{\em JHEP} {\bfseries
  12} (2001) 023}, \href{http://arxiv.org/abs/hep-th/0111177}{{\ttfamily
  arXiv:hep-th/0111177}}.

\bibitem{Melvin:1963qx}
M.~A. Melvin, ``{Pure magnetic and electric geons},''
  \href{http://dx.doi.org/10.1016/0031-9163(64)90801-7}{{\em Phys. Lett.}
  {\bfseries 8} (1964) 65--70}.

\bibitem{Debray:2021vob}
A.~Debray, M.~Dierigl, J.~J. Heckman, and M.~Montero, ``{The anomaly that was
  not meant IIB},'' \href{http://arxiv.org/abs/2107.14227}{{\ttfamily
  arXiv:2107.14227 [hep-th]}}.

\bibitem{deBoer:2001wca}
J.~de~Boer, R.~Dijkgraaf, K.~Hori, A.~Keurentjes, J.~Morgan, D.~R. Morrison,
  and S.~Sethi, ``{Triples, fluxes, and strings},''
  \href{http://dx.doi.org/10.4310/ATMP.2000.v4.n5.a1}{{\em Adv. Theor. Math.
  Phys.} {\bfseries 4} (2002) 995--1186},
  \href{http://arxiv.org/abs/hep-th/0103170}{{\ttfamily arXiv:hep-th/0103170}}.

\bibitem{Tachikawa:2015wka}
Y.~Tachikawa, ``{Frozen singularities in M and F theory},''
  \href{http://dx.doi.org/10.1007/JHEP06(2016)128}{{\em JHEP} {\bfseries 06}
  (2016) 128}, \href{http://arxiv.org/abs/1508.06679}{{\ttfamily
  arXiv:1508.06679 [hep-th]}}.

\bibitem{FrozenToAppear}
M.~Cvetic, M.~Dierigl, L.~Lin, E.~Torres, and H.~Y.~Zhang {\em {\textit{To
  Appear}}} .

\bibitem{Braeger:2024jcj}
N.~Braeger, V.~Chakrabhavi, J.~J. Heckman, and M.~Hubner, ``{Generalized
  Symmetries of Non-Supersymmetric Orbifolds},''
  \href{http://arxiv.org/abs/2404.17639}{{\ttfamily arXiv:2404.17639
  [hep-th]}}.

\bibitem{Cvetic:2023pgm}
M.~Cveti\v{c}, J.~J. Heckman, M.~H\"ubner, and E.~Torres, ``{Generalized
  symmetries, gravity, and the swampland},''
  \href{http://dx.doi.org/10.1103/PhysRevD.109.026012}{{\em Phys. Rev. D}
  {\bfseries 109} no.~2, (2024) 026012},
  \href{http://arxiv.org/abs/2307.13027}{{\ttfamily arXiv:2307.13027
  [hep-th]}}.

\bibitem{Gould:2023wgl}
D.~S.~W. Gould, L.~Lin, and E.~Sabag, ``{Swampland Constraints on the SymTFT of
  Supergravity},'' \href{http://arxiv.org/abs/2312.02131}{{\ttfamily
  arXiv:2312.02131 [hep-th]}}.

\bibitem{Cvetic:2022uuu}
M.~Cveti\v{c}, M.~Dierigl, L.~Lin, and H.~Y. Zhang, ``{All eight- and
  nine-dimensional string vacua from junctions},''
  \href{http://dx.doi.org/10.1103/PhysRevD.106.026007}{{\em Phys. Rev. D}
  {\bfseries 106} no.~2, (2022) 026007},
  \href{http://arxiv.org/abs/2203.03644}{{\ttfamily arXiv:2203.03644
  [hep-th]}}.

\end{thebibliography}\endgroup

\end{document}